\documentclass[cameraready]{Interspeech}

\usepackage{booktabs, multirow, xcolor, array}
\newcommand{\xmark}{\textcolor{red!70!black}{\ding{55}}}
\usepackage{pifont}
\usepackage{times}
\usepackage{latexsym}
\usepackage[T1]{fontenc}
\usepackage[utf8]{inputenc}
\usepackage{microtype}
\usepackage{inconsolata}
\usepackage{graphicx}
\usepackage{booktabs}
\usepackage{amssymb}
\usepackage{hyperref} 
\usepackage{multirow}
\usepackage{arydshln}
\usepackage{cite}
\usepackage{xcolor}
\usepackage{listings}
\usepackage[most]{tcolorbox}
\usepackage{pifont}
\usepackage{makecell}
\usepackage[table]{xcolor}
\usepackage{colortbl}
\newcommand{\cmark}{\ding{51}}
\definecolor{lightgray}{gray}{0.92}
\tcbuselibrary{listings,skins,breakable}

\usepackage[precision=2, unit=mm]{lengthconvert}

\tcbuselibrary{listings,skins,breakable}

\title{ESPnet3: Infrastructure for Scalable Speech and Audio Research\\in the Foundation Model Era}

\author[affiliation={1}, orcid=0009-0006-5864-7634]{Masao}{Someki}
\author[affiliation={2}, orcid=0009-0000-4958-202X]{Alexander}{Polok}
\author[affiliation={3}, orcid=0000-0002-9603-2423]{Carlos}{Carvalho}
\author[affiliation={1}, orcid=0000-0002-2447-9330]{Chyi-Jiunn}{Lin}
\author[affiliation={1,4}, orcid=0009-0008-6253-2543]{Da-Hee}{Yang}
\author[affiliation={1}, orcid=0000-0002-9050-8304]{Jiatong}{Shi}
\author[affiliation={1}, orcid=0009-0002-5461-4342]{Jinchuan}{Tian}
\author[affiliation={5}, orcid=0000-0001-7907-0076]{Nelson Enrique}{Yalta Soplin}
\author[affiliation={1}, orcid=0000-0002-5358-1844]{Samuele}{Cornell}
\author[affiliation={1}, orcid=0000-0003-0375-496X]{Siddhant}{Arora}
\author[affiliation={3}, orcid=0000-0001-6614-3168]{Francisco}{Teixeira}
\author[affiliation={6}, orcid=0009-0006-9498-1592]{Wei}{Wang}
\author[affiliation={1}, orcid=0000-0002-3251-3084]{William}{Chen}
\author[affiliation={3}, orcid=0000-0000-0000-0000]{Alberto}{Abad}
\author[affiliation={6}, orcid=0000-0003-0299-9914]{Chenda}{Li}
\author[affiliation={1}, orcid=0000-0002-5970-8631]{Shinji}{Watanabe}
\author[affiliation={6}, orcid=0000-0003-4500-3515]{Wangyou}{Zhang}

\makeatletter
\g@addto@macro\authorlist{\\[-1mm]\textnormal{\small\bfseries (Authors are listed in alphabetical order.)}}
\makeatother

\address{
    $^1$ Pittsburgh, USA Carnegie Mellon University
    $^2$ Brno, Czechia Brno University of Technology \\
    $^3$ Lisbon, Portugal Instituto Superior Técnico
    $^4$ Seoul, South Korea Hanyang University \\
    $^5$ Tokyo, Japan Hitachi Astemo 
    $^6$ Shanghai, China Shanghai Jiao Tong University
}

\email{msomeki@andrew.cmu.edu}


\usepackage{comment}

\begin{document}

\maketitle

\begin{abstract}
Recent speech research involves increasingly large datasets, complex models, and diverse experimental workflows. However, existing frameworks require substantial engineering effort to support such experiments. We present ESPnet3, a speech and audio research framework built on a modular system architecture with configuration-driven dataset composition and unified Python-based workflows. ESPnet3 introduces a DataOrganizer abstraction for flexible dataset integration and dataset sharding for memory-efficient large-scale training, while allowing recipe-specific logic through lightweight stage overrides. In OWSM pre-training experiments, ESPnet3 reduces per-epoch training time by \emph{21.1 minutes} compared to ESPnet2 and achieves \emph{>80\% GPU utilization} in multi-node training. Fine-tuning experiments show that new models and datasets can be integrated with around \emph{46 lines of additional code}. ESPnet3 will be publicly released with model checkpoints and training logs.

\end{abstract}


\section{Introduction}
Speech and audio research has entered the foundation-model era.
Training corpora have expanded from thousands to millions of hours~\cite{whisper, google_usm, moshi, SeamlessM4T, phi4, Qwen3-ASR, owsm},
while model sizes have grown to billions of parameters~\cite{yue,granite,Qwen-Audio,kimi-audio, audio-flamingo, step_audio2, owls, bagpiper}.
Large-scale training now enables unified models that cover diverse languages, accents, and tasks within a single architecture~\cite{google_usm,mms,xeus}.
At the same time, research is increasingly focused on systems that combine multiple pretrained models, external tools, and multi-task pipelines~\cite{espnet-sds,deng2023university, notsofar,module_far, MITROFANOV2025101780, kamo24_chime, ye2023iacas, wangustc}. 
This transition has fundamentally changed not only model capacity, but also the requirements placed on research infrastructure.

Modern deep-learning-based toolkits such as ESPnet~\cite{espnet}, Fairseq~\cite{ott-etal-2019-fairseq}, and HuggingFace Transformers~\cite{wolf-etal-2020-transformers} have significantly lowered the entry barrier for speech research.
However, as experiments scale to multi-million-hour corpora and billion-parameter models, system-level challenges emerge that extend beyond model implementation.
These include composing multiple datasets, efficient dataset iteration, distributed training across GPUs and nodes, and integrating large pretrained models with parameter-efficient fine-tuning (PEFT) methods.

ESPnet and its underlying framework ESPnet2 have played a central role in the open-source speech community by providing unified implementations across a wide range of tasks.
However, its architecture tightly couples the experiment logic with the core components, which makes new experimental configurations costly to implement.
For example, adding LoRA~\cite{lora} support for Whisper~\cite{whisper} required modifications across more than 20 files and 670 lines of code in ESPnet2\footnote{Including CI and test scripts, the PR \href{https://github.com/espnet/espnet/pull/5400}{\#5400} modified 26 files and 899 lines.}.
ESPnet-EZ~\cite{espnetez} was introduced to address some of these usability concerns
by enabling Python-only interaction and simplifying fine-tuning workflows.
However, it was not designed for large-scale training from scratch and lacks abstractions for data handling or distributed training.

To address these challenges, we propose \emph{ESPnet3}, a redesigned speech and audio research framework that builds upon ESPnet2's role in enabling large-scale research while reducing the recipe-level engineering effort required for exploratory experimentation.
ESPnet3 adopts a modular system architecture in which the common stages of the research workflow are centralized within the framework, while experiment-specific logic remains in lightweight recipes.
This separation reduces coupling between experiments and the internals of the framework
while maintaining compatibility with ESPnet models such as Conformer~\cite{conformer} and E-Branchformer~\cite{e_branchformer}.
We evaluate ESPnet3 using large-scale OWSM~\cite{owsm} pre-training and fine-tuning experiments.
In pre-training experiments, ESPnet3 reduces per-epoch training time by 21.1 minutes (22.2\% relative) compared to ESPnet2 while achieving stable multi-node GPU utilization above 80\%.
Fine-tuning experiments further show that new models and datasets can be integrated with minimal additional code.

The main contributions of this work are as follows:
\begin{itemize}
    \item \textbf{Configuration-driven data abstraction:} A Hydra~\cite{hydra}-based \texttt{DataOrganizer} that enables modular dataset composition and shard-based iteration for scalable training.
    \item \textbf{Modular system architecture:} A modular design that separates experiment logic from core infrastructure, reducing architectural coupling, and improving extensibility.
    \item \textbf{End-to-end workflow:}
    A unified framework that connects data processing, training, evaluation, inference, and publication, integrating evaluation platforms such as VERSA~\cite{shi2025versa}.
    \item \textbf{Scalable experimentation support:} Support for large-scale pre-training and third-party PEFT libraries within a unified workflow, enabling new models and datasets to be incorporated with minimal additional code.
\end{itemize}

\section{Related Works}

\begin{table}[t]
\centering
\caption{Task and feature coverage across speech processing frameworks. ESPnet3 inherits task coverage from ESPnet2 while introducing new infrastructure for scalable experimentation.
Frameworks:
 E2=ESPnet2, 
 E3=ESPnet3,
 Nemo, 
 SB=SpeechBrain, 
 NGK=Next-gen Kaldi, 
 ST=Speech Translation, SLU=Spoken Language Understanding.
\cmark{} = supported, \xmark{} = not supported.}
\label{tab:coverage}
\footnotesize
\setlength{\tabcolsep}{4pt}
\begin{tabular}{@{}l ccccc@{}}
\toprule
 & \textbf{E2} & \textbf{E3 (Ours)} & \textbf{NeMo} & \textbf{SB} & \textbf{NGK} \\
\midrule
\multicolumn{6}{@{}l}{\textit{Speech and Audio Modeling Tasks}} \\
\quad ASR                    & \cmark & \cmark & \cmark & \cmark & \cmark \\
\quad ST     & \cmark & \cmark & \cmark & \cmark & \xmark \\
\quad SLU    & \cmark & \cmark & \cmark & \cmark & \xmark \\
\quad TTS                    & \cmark & \cmark & \cmark & \cmark & \cmark \\
\quad Voice Conversion       & \cmark & \cmark & \xmark & \xmark & \xmark \\
\quad Singing Voice Synth.   & \cmark & \cmark & \xmark & \xmark & \xmark \\
\quad Speech Enhancement     & \cmark & \cmark & \cmark & \cmark & \xmark \\
\quad Speaker Recognition    & \cmark & \cmark & \cmark & \cmark & \cmark \\
\quad Self-supervised Learning & \cmark & \cmark & \cmark & \cmark & \xmark \\
\quad Speech-language Models   & \cmark & \cmark & \cmark & \cmark & \xmark \\
\quad Audio Classification     & \cmark & \cmark & \cmark & \cmark &  \cmark \\
\quad Neural Speech Codec      & \cmark & \cmark & \cmark & \xmark & \cmark \\
\midrule
\multicolumn{6}{@{}l}{\textit{Infrastructure}} \\
\quad Multi-node Training      & \cmark & \cmark & \cmark & \cmark & \cmark \\
\quad HF Datasets Integration  & \xmark & \cmark & \cmark & \cmark & \cmark \\
\quad Dataset Sharding         & \xmark & \cmark & \cmark & \xmark & \cmark \\
\quad Pure Python Workflow     & \xmark & \cmark & \cmark & \cmark & \xmark \\
\bottomrule
\end{tabular}
\end{table}
Several open-source toolkits~\cite{kaldi,espnetez} have been developed for end-to-end speech and audio processing, each emphasizing different design priorities and user communities.
These toolkits differ in both task coverage and infrastructure design, as summarized in Table~\ref{tab:coverage}.
ESPnet~\cite{espnet} supports a broad range of speech modeling tasks, while other frameworks prioritize different subsets of these capabilities.
With respect to scalability, modern frameworks increasingly support distributed training and large pretrained models.
NeMo~\cite{nemo} explicitly targets large-scale and multi-node training scenarios, while SpeechBrain~\cite{speechbrain,speechbrain_v1} and other PyTorch-based frameworks rely on native distributed backends.
ESPnet3 builds on the ESPnet ecosystem and focuses on restructuring the underlying research infrastructure to better support large-scale experimentation.
Specifically, ESPnet3 introduces configuration-driven dataset composition, dataset sharding for memory-efficient training, and a modular pythonic workflow that separates experiment logic from core framework components.

By adopting Hydra-based configuration and PyTorch Lightning–based training, ESPnet3 aligns ESPnet workflows with widely used modern deep-learning infrastructure
while remaining compatible with existing ESPnet2-based recipes.
Together, these dependencies improve long-term maintainability and allow ESPnet3 to support large datasets, distributed training, and parameter-efficient fine-tuning within a unified workflow.

\begin{figure}[t]
  \centering
  \begin{minipage}[t]{0.47\linewidth}
    \centering
    \includegraphics[width=\linewidth]{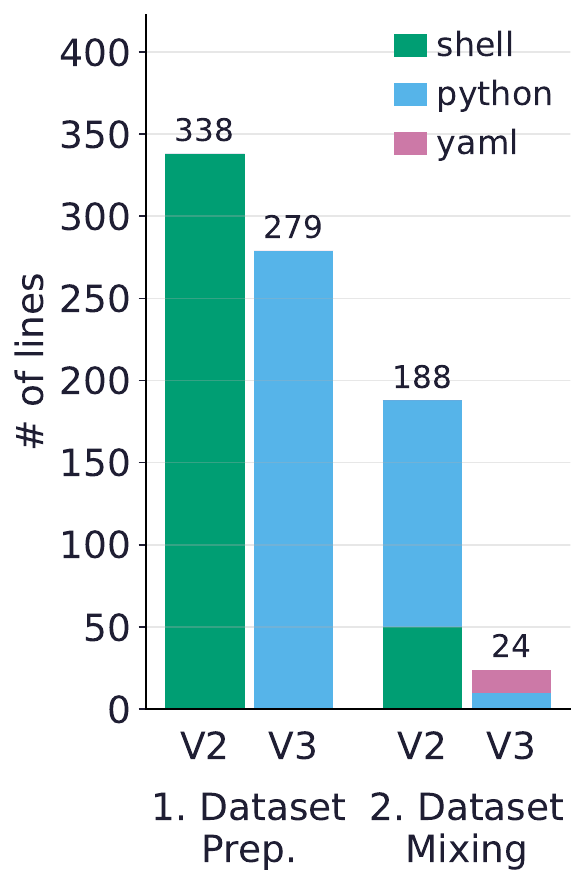}
  \end{minipage}\hfill
  \begin{minipage}[t]{0.47\linewidth}
    \centering
    \includegraphics[width=\linewidth]{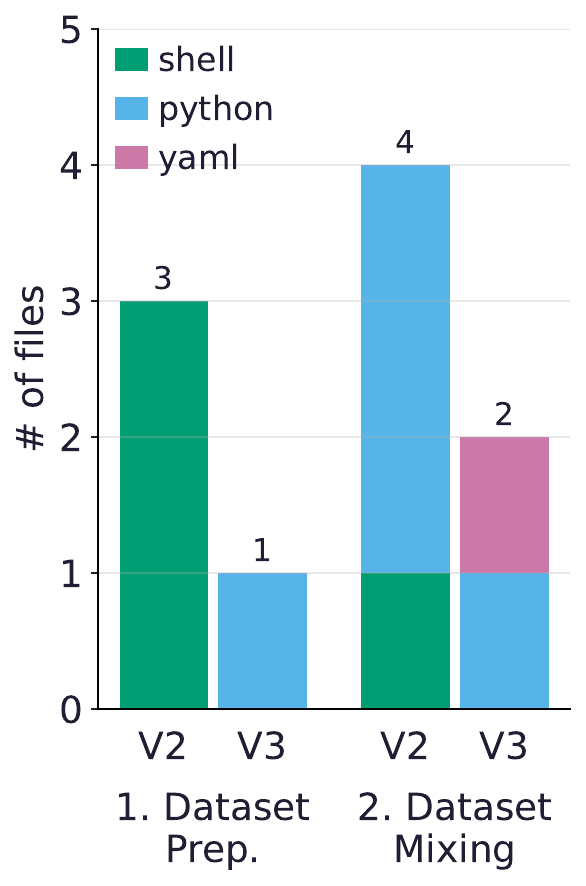}
  \end{minipage}

  \caption{
  Comparison of implementation effort between ESPnet2 (V2) and ESPnet3 (V3) for dataset management.
  (a) ESPnet3 significantly reduces the lines of code to prepare and mix the datasets for large scale training.
  (b) The number of files required to edit is also minimized.
  }
  \label{fig:number-of-lines}
\end{figure}

\section{Design}
\begin{figure}[t]
  \begin{minipage}{\linewidth}
    \raggedleft
    \includegraphics[width=0.48\linewidth]{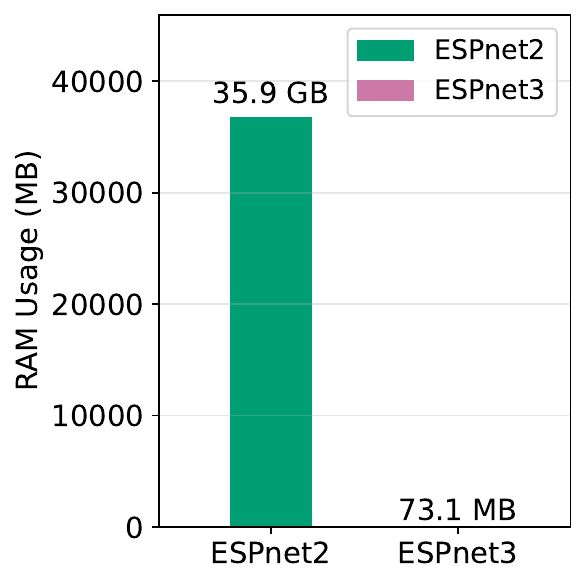}%
    \hfill
    \includegraphics[width=0.48\linewidth]{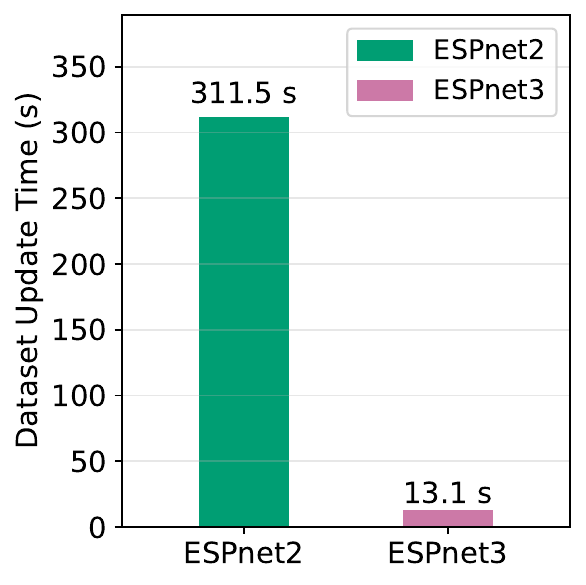}
  \end{minipage}
  \caption{Memory overhead (left) and dataset refresh time (right) in
our OWSM pre-training experiments: ESPnet2 without dataset sharding vs. ESPnet3 with sharding. (64 shards, 16 GPUs).}
  \label{fig:pretrain_metrics}
\end{figure}

ESPnet3 is designed around three architectural principles: configuration-driven data abstraction for declarative dataset composition and scalable iteration, modular system architecture that separates experiment logic from framework internals, and a unified end-to-end workflow that connects data processing, training, inference, and evaluation.




\subsection{Data Organizer}
\label{sec:data_organizer}
Large-scale speech projects often require integrating dozens of heterogeneous corpora, each with distinct formats and preprocessing pipelines. 
In ESPnet2-based workflows, such integration frequently relied on recipe-level scripting and format conversion, leading to extensive glue code when datasets were added, mixed, or augmented. 
These challenges became particularly evident in large-scale projects such as OWSM~\cite{owsm} and Bagpiper~\cite{bagpiper}, which motivated a structural redesign of data handling within the ESPnet framework.
ESPnet3 introduces \texttt{DataOrganizer}, a configuration-driven abstraction that treats datasets as modular components. 
Datasets are declaratively composed through Hydra-based YAML configuration:
\begin{tcolorbox}[
  fontupper=\footnotesize,
  left=2mm,
  width=\linewidth
]
\begin{lstlisting}
dataset:
  _target_: DataOrganizer
  train:
    - dataset: Librispeech
    - dataset: Switchboard
\end{lstlisting}
\end{tcolorbox}
Additional datasets can be enabled or disabled by editing the configuration, without modifying pipeline code. 
Each dataset is implemented as a standard PyTorch \texttt{Dataset} returning a dictionary of fields, and the \texttt{DataOrganizer} unifies them into a single iterable object while preserving dataset-specific logic.
Figure~\ref{fig:number-of-lines} quantifies the resulting reduction in engineering effort. 
Compared to ESPnet2, ESPnet3 significantly reduces both lines of code and the number of files required for dataset preparation and mixing.

\subsection{Dataset Sharding}
\label{sec:shard}
Full-split dataset iteration becomes inefficient at the million-hour scale due to repeated initialization and memory overhead.
ESPnet3 addresses this through shard-level iteration.
Let $S$ denote the number of shards and $R$ the number of distributed workers.
At epoch $e$, worker $r$ processes shard
\begin{equation}
s(e,r) = (eR + r) \bmod S.
\end{equation}
This rank-aware shard rotation ensures balanced coverage without reloading the full dataset at epoch boundaries.

As shown in Figure~\ref{fig:pretrain_metrics}, shard-based iteration significantly reduces dataset initialization overhead in our OWSM pre-training experiments (64 shards, 16 GPUs).
For ESPnet2, we report the total size of dataset metadata, which must be loaded on RAM during DataLoader construction, while for ESPnet3 we report the measured increase in CPU RAM during shard-level DataLoader initialization.
ESPnet2 requires materializing 36\,GB dataset metadata, whereas ESPnet3 leverages lazy loading via HuggingFace Datasets and requires only 73\,MB of additional CPU RAM usage on average.
Dataset refresh time is also reduced from 311.5\,s to 13.1\,s due to the reduced memory overhead.

\subsection{Recipe Overview}
\label{sec:recipe_overview}

\begin{figure}[t]
  \includegraphics[width=0.9\linewidth]{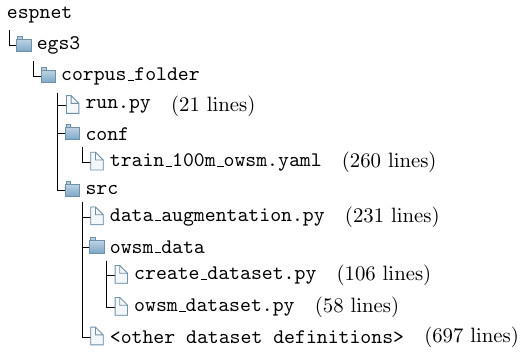}
  \caption{Typical directory structure of an ESPnet3 training recipe,
illustrated using the OWSM pretraining setup. Numbers in parentheses
indicate lines of code.}
  \label{fig:owsm-recipe}
\end{figure}

\begin{figure}[t]
    \centering
    \includegraphics[width=\linewidth]{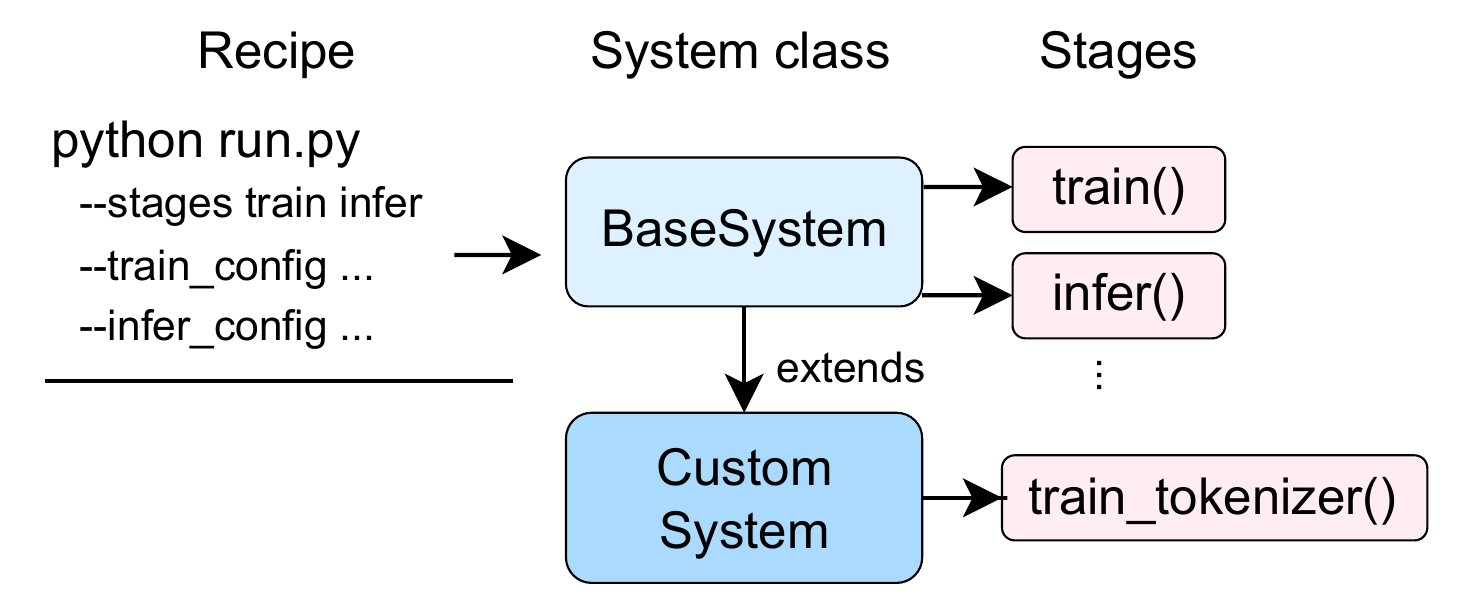}
    \caption{
Execution architecture of ESPnet3.
The \texttt{run.py} entry point loads experiment configurations
and initializes the \texttt{System} class, which implements
workflow stages such as training and inference.
Recipe-specific functionality can be introduced by overriding
\texttt{BaseSystem} and adding stage functions.
}
    \label{fig:system_arch}
\end{figure}

Figure~\ref{fig:owsm-recipe} illustrates a typical directory structure of an ESPnet3 training recipe using the OWSM pretraining setup.
In ESPnet3, \texttt{run.py} serves as the entry point of the workflow, loading experiment configurations defined in YAML and executing the corresponding stages.
In ESPnet2, execution stages were defined within task-specific shell scripts, and their definitions often differed across recipes.
This led to duplicated logic and inconsistencies across tasks, since improvements to training pipelines or evaluation procedures had to be manually propagated to each recipe.
ESPnet3 addresses this issue by centralizing common workflow stages within the core framework through a shared \texttt{BaseSystem} abstraction. The \texttt{BaseSystem} class serves as a centralized orchestration layer that manages common workflow stages, including data preparation, training, inference, and model packaging, as well as optional evaluation/measurement stages when configured.
As shown in Figure~\ref{fig:system_arch}, the system provides standard stages such as \texttt{train()} and \texttt{infer()}, ensuring that updates to training logic automatically apply across tasks.
Recipe-specific functionality can be introduced by extending \texttt{BaseSystem} and implementing additional stage functions.
Instead of large shell and Perl orchestration scripts used in ESPnet2, ESPnet3 recipes consist of a lightweight Python entry point, configuration files, and modular source components.
As a result, the OWSM training recipe shrinks from 2{,}289 lines of orchestration code in ESPnet2 to just 70 lines in ESPnet3 (Table~\ref{tab:efficiency}).
\section{Experiments}

This section evaluates ESPnet3 from a system perspective, focusing on implementation simplicity and extensibility rather than raw model performance.
We use OWSM V4 pre-training and fine-tuning as representative case studies to demonstrate how ESPnet3 supports multiple dimensions of scaling, including large-scale dataset integration and diverse fine-tuning strategies.

\subsection{OWSM Pre-training}

\textbf{Setup. }
We conducted pre-training of the OWSM-V4 base model~\cite{owsm} (102\,M parameters) to evaluate both the implementation simplicity and the effectiveness of dataset scaling in ESPnet3.
This experiment leverages the \texttt{DataOrganizer} and 
sharding mechanisms described in Sections~\ref{sec:data_organizer} and 
\ref{sec:shard}.
The model was trained on approximately 320\,k hours of speech data.
To further demonstrate the flexibility of ESPnet3's data pipeline, we applied on-the-fly data augmentation during training following~\cite{granite}, which can be seamlessly integrated via the \texttt{DataOrganizer} without modifying the training pipeline.
%
Notably, this pre-training was carried out using the default
\texttt{BaseSystem} provided by ESPnet3 (Section~\ref{sec:recipe_overview}), without introducing any
OWSM-specific system implementations or modifying the core framework.
Word error rate (WER) was evaluated on the CHiME-4~\cite{chime4} real test set.


\begin{figure}[t]
  \centering
  \includegraphics[width=\linewidth]{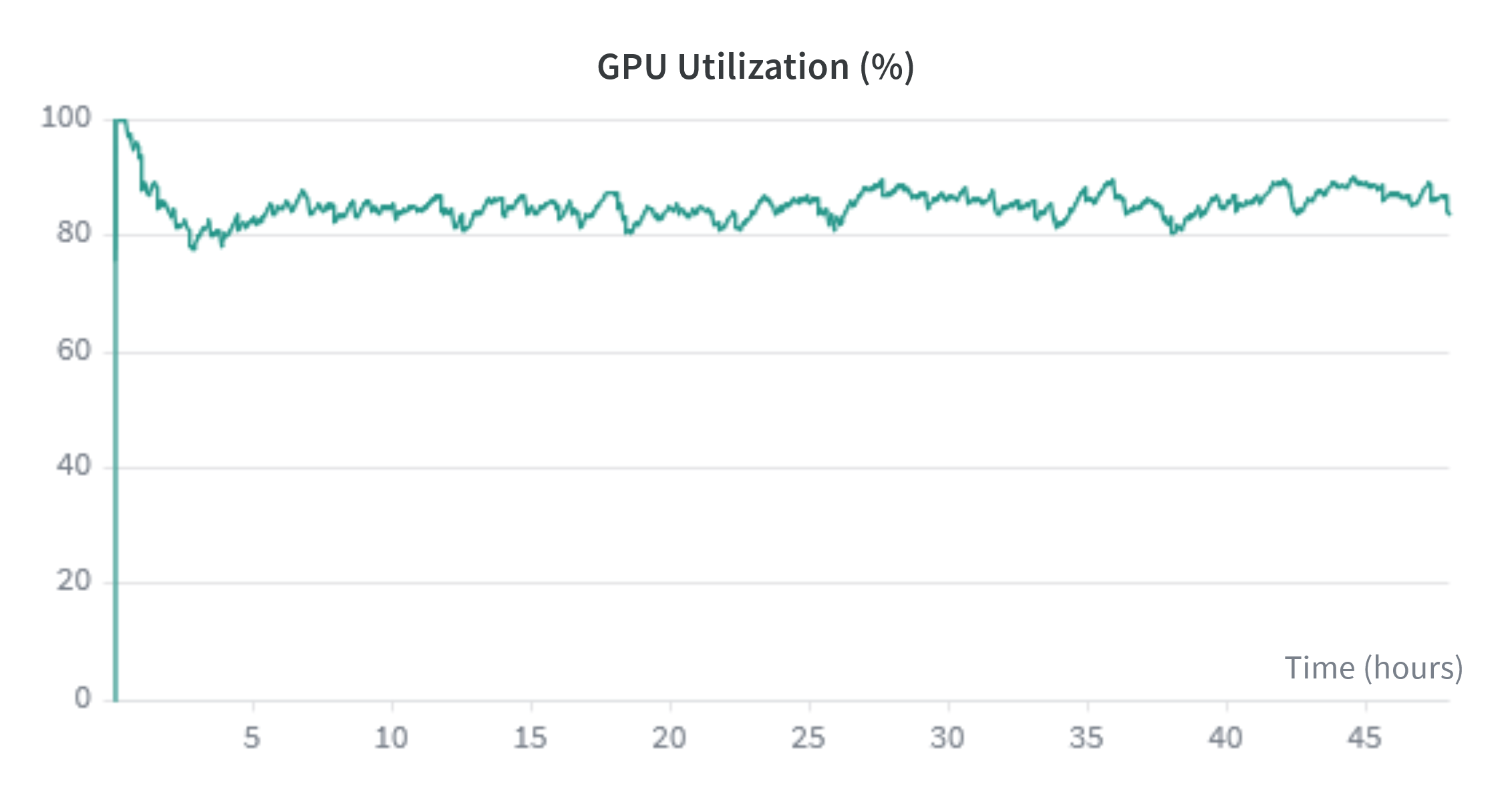}
  \caption{
  GPU utilization for 4-node (16 GPUs, accumulation = 1) training.
We achieved over 80\% scaling efficiency even across nodes (Slingshot interconnect).
}
  \label{fig:gpu_utilization}
\end{figure}

\noindent\textbf{Results.}
All reported timing results in this section are averaged over logs from five consecutive training epochs.
As shown in Figure~\ref{fig:gpu_utilization}, OWSM pre-training on ESPnet3 achieves stable GPU utilization exceeding 80\% on a 4-node, 16-GPU setup (4 H100 per node).
Table~\ref{tab:efficiency} compares the training efficiency between
ESPnet2 and ESPnet3 for OWSM-Base pre-training.
Compared to the ESPnet2 baseline\footnote{\href{https://huggingface.co/espnet/owsm_v4_base_102M}{espnet/owsm\_v4\_base\_102M}},
ESPnet3 reduces the average epoch time from 95.3 minutes to 74.2 minutes,
and the per-update processing time from 0.594 seconds to 0.441 seconds.

\begin{table}[t]
\centering
\caption{
Comparison of training efficiency and implementation complexity
between ESPnet2 and ESPnet3 for OWSM-Base pre-training.
Both experiments were conducted with the same hardware and training configurations.
\textbf{Upd} denotes the average wall-clock time per optimizer update (seconds), and the results are averaged over 5 epochs.
\textbf{Config}, \textbf{Dataset}, and \textbf{Recipe} indicate the number of lines of code.
}
\label{tab:efficiency}
\resizebox{\linewidth}{!}{
\begin{tabular}{lccccc}
\toprule
Framework & Upd (s) & Epoch (min) & Config & Dataset & Recipe \\
\midrule
ESPnet2 & 0.594 & 95.3 & 147 & 1230 & 2289 \\
ESPnet3 & 0.441 & 74.2 & 268 & 1723 & 70 \\
\bottomrule
\end{tabular}
}
\end{table}
Furthermore, ESPnet3 reduces the overhead of large-scale dataset
iteration.
In ESPnet2, handling large datasets often requires periodically reconstructing the data iterator to avoid excessive memory usage.
In contrast, ESPnet3 uses dataset sharding (Section~\ref{sec:shard}), which reduces RAM usage from around 35.9GB to 73.1MB and dataset refresh time from 311.5s to 13.1s, as shown in Figure~\ref{fig:pretrain_metrics}.

\begin{table}[t]
\centering
\caption{
Effect of data augmentation in OWSM-Base pre-training.
WER is evaluated on the CHiME-4 test set at the same training budget (350k updates).
}
\label{tab:augmentation}
\begin{tabular}{lcc}
\toprule
Setting & WER (\%) & $\Delta$WER \\
\midrule
ESPnet3 (plain) & 12.84 & -- \\
ESPnet3 (+ augmentation) & 12.53 & -0.31 \\
\bottomrule
\end{tabular}
\end{table}
Table~\ref{tab:augmentation} shows that enabling augmentation through the \texttt{DataOrganizer} pipeline leads to an improvement in WER, confirming that standard training workflows can be seamlessly integrated.

\subsection{Fine-tuning Results}
\textbf{Setup. }
We also conducted fine-tuning experiments with 
Whisper Large v3 (WhisperLv3)~\cite{whisper} to demonstrate that ESPnet3 supports large-scale custom model training, even for models that are not natively part of the ESPnet framework.
More specifically, we fine-tuned the 
WhisperLv3 on FalAR~\cite{falar}, a large-scale European Portuguese (EP) parliamentary speech dataset derived from the Portuguese National Assembly, which includes 5k hours of speaker-annotated data recorded over a span of 20 years.
For the FalAR setup, we evaluated the fine-tuned models on the CAMÕES benchmark, a multi-domain collection of 14 datasets, spanning a diverse set of speakers, with ages ranging from 3-to-100 years, designed to benchmark modern EP ASR systems~\cite{camoes}.
To further demonstrate the ease of model customization, we also performed PEFT using LoRA via the HuggingFace \texttt{PEFT} package.
For inference with all models, we use a beam size of 1 and apply a Portuguese-specific normalizer derived from Whisper's base normalizer.


\begin{table}[!t]\centering
\caption{
WER (\%) comparison of
Whisper Large v3 (WhisperLv3): zero-shot results on CAMÕES benchmark and fine-tuned results after training both models on FalAR dataset using ESPnet3.
(OWSM results will be added in the future release)
 }
\label{tab:finetune}
\resizebox{\linewidth}{!}{ 
\begin{tabular}{lcccccc}\toprule
\multicolumn{3}{c}{model} &WER (\%) &\#lines & \#files \\\midrule
\multicolumn{3}{c}{WhisperLv3} &22.65 &- &- \\\midrule
\multicolumn{6}{c}{\textbf{Finetuned on FalAR}} \\
&\multirow{2}{*}{WhisperLv3} &full finetune & 19.42& 212 & 5\\
& &+ Lora (PEFT) &19.47 & 297 & 5\\
\bottomrule
\end{tabular}
}
\end{table}
\noindent\textbf{Results. }
As shown in Table~\ref{tab:finetune}, ESPnet3 enables fine-tuning
for models that are not originally supported by ESPnet, such as Whisper.
Furthermore, integrating a new Hugging Face dataset requires only 46 lines of code in ESPnet3.
For comparison, implementing the same functionality in ESPnet2 required 374 lines of code in our manual implementation, corresponding to an 87.7\% relative reduction.
These experiments show the potential of ESPnet3 in delivering state-of-the-art models, trained with large-scale data, with minimal additional code. 

\section{Conclusion}

In this paper, we present ESPnet3, a redesigned speech and audio research framework that addresses system-level bottlenecks that emerge in large-scale speech modeling.
As speech and audio research scales in data, model, language, and research styles, engineering overhead has become a critical constraint on research productivity.
ESPnet3 adopts a modular and configuration-driven design that separates system logic, data organization, and experimental recipes.
As a result, ESPnet3 significantly reduces the required development in the recipe-level code size compared to ESPnet2.
In large-scale OWSM pre-training, ESPnet3 shows a reduction of the average per-epoch training time by approximately 21.1 minutes, showing efficiency in multi-node training.
Experiments on fine-tuning the
Whisper model demonstrate that ESPnet3 enables seamless integration of external and custom models, supporting diverse training and fine-tuning strategies within a unified, scalable workflow.
By lowering the entry barrier to speech and audio modality development, we expect ESPnet3 to encourage researchers to explore multimodal systems that integrate with other modalities.
\section{Generative AI Use Disclosure}
\label{sec:ai_usage}


Generative AI tools have been used to help revise and refine the manuscript. 
Authors take full responsibility for the content.

\section{Acknowledgements}
Experiments of this work used the Bridges2 system at PSC and Delta and DeltaAI system at NCSA through allocations CIS210014 and IRI120008P from the Advanced Cyberinfrastructure Coordination Ecosystem: Services \& Support (ACCESS) program, supported by National Science Foundation grants \#2138259,\#:2138286, \#:2138307, \#:2137603, and \#:2138296.

The work was also partially supported by Ministry of Education, Youth and Sports of the Czech Republic (MoE) through the OP JAK project "Linguistics, Artificial Intelligence and Language and Speech Technologies: from Research to Applications" (ID:CZ.02.01.01/00/23\_020/0008518).
\bibliographystyle{IEEEtran}
\bibliography{mybib}

\end{document}